# PLANET SIZE DISTRIBUTION FROM THE KEPLER MISSION AND ITS IMPLICATIONS FOR PLANET FORMATION.

Li Zeng[1], Stein B. Jacobsen[1], Eugenia Hyung[1], Andrew Vanderburg[2], Mercedes Lopez-Morales[2], Dimitar D. Sasselov[2], Juan Perez-Mercader[1], Michail I. Petaev[1,2], David W. Latham[2], Raphaëlle D. Haywood[2], and Thomas K. R. Mattson[3]. [1]Department of Earth & Planetary Sciences, Harvard University, 20 Oxford Street, Cambridge, MA 02138 (astrozeng@gmail.com), [2]Harvard-Smithsonian Center for Astrophysics, 60 Garden Street, Cambridge, MA 02138, [3]Sandia National Laboratories, PO Box 5800 MS 1189, Albuquerque, NM 87185.

**Introduction:** The overall size distribution of exoplanets found by the Kepler mission so far appear to best fit a log-normal distribution, indicative of a population formed by a material-limited and time-limited growth process. Further detailed analysis of their size distribution with respect to that of the stellar fluxes they receive suggests a bimodal modification on top of the log-normal distribution - a division of planets into two groups: rocky planets (<2 $R_\oplus$) and water-rich planets (>2 $R_\oplus$) with or without gaseous envelopes.

**Planet Size Distribution from Observations:**

*Log-Normal Distribution.* Analyzing the overall radius distribution of 5000+ Kepler planet candidates [1], we have obtained the best fit to a log-normal distribution, indicative of growth process, as planets are formed by growing from small bits and pieces. In nature, things that grow from small increments where each increment is stochastic follows a log-normal distribution (growth rate proportional to a certain power of its size or mass), such as the size distributions of apples, our finger nails, and pumpkins (yes, pumpkins, as occasionally we find pumpkins that grown really big, but those are rare), and the size distribution of cities on Earth [2].

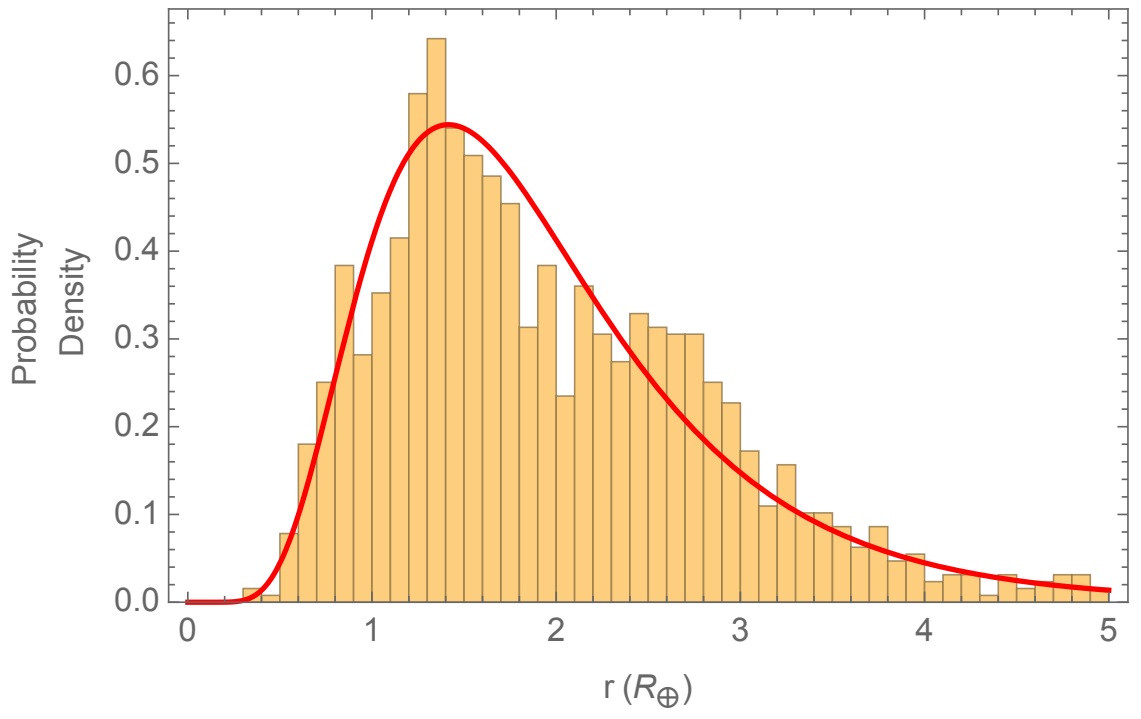

**Figure 1.** One-dimensional histogram of planet radii.

Our galaxy is an orchard, in which stars and planets grow. They grow from small bits and pieces, thus obeying a log-normal distribution. Log-normal distribution maximizes the entropy, implying that the growth process is stochastic. Anything that grows by accumulation of random bits and pieces, with each increment stochastic, one would get a log-normal distribution. There is a completeness correction that may need to be applied to planets <~1.5 $R_\oplus$ [3].

*Bi-Modal Distribution.* More careful analysis of plotting 2-D histogram of fluxes received Vs. radii reveals a bi-modal distribution on top of the log-normal. The modes are centered around 1.3 $R_\oplus$ and 2.3 $R_\oplus$, with a gap at 2 $R_\oplus$. This seems to be a universal feature throughout FGKM stars. These two modes could correspond to two types of planets in general, independent of the host star type (especially taking into account that M-Dwarf radius in Kepler Input Catalog (KIC) could be underestimated by ~20% [4]).

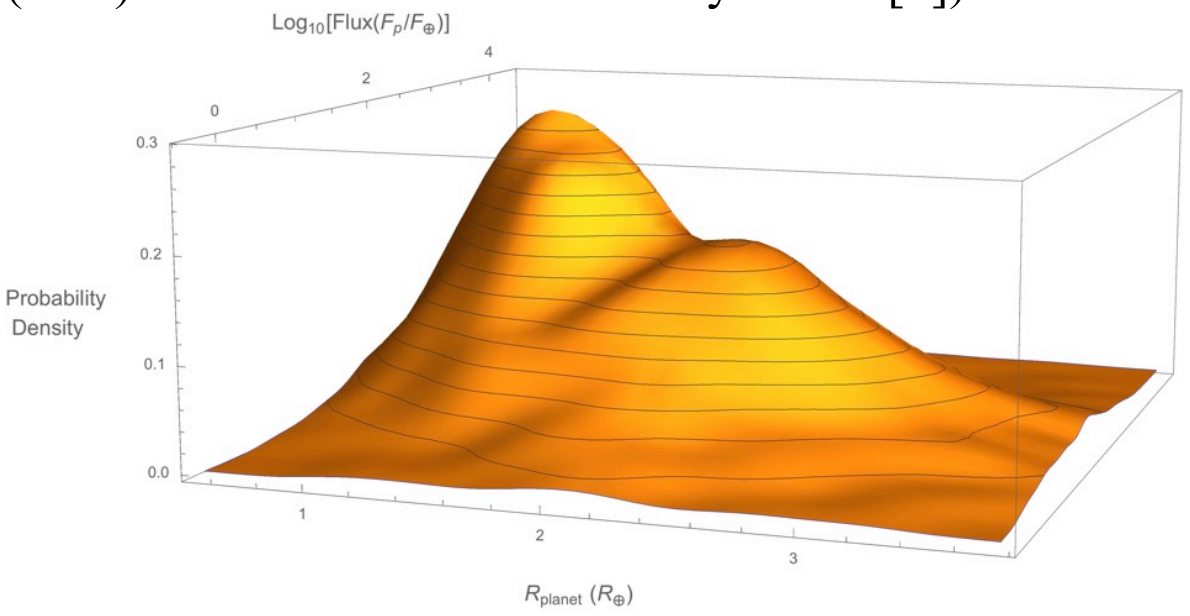

**Figure 2.** Two-dimensional smoothed histogram of planet radii versus stellar fluxes they receive.

**Interpretations:**

*Two types of Planets: rocky and water worlds.* These two distinct types of planets are rocky ones (Earths and super-Earths, planets made up of mostly silicates and metals with bulk composition similar to Earth), and volatile-rich ones. The volatile-rich planets should be water-worlds (made of a significant amount (>25%) of H-compounds: $H_2O$, $NH_3$, $CH_4$, in addition to the silicates/metal and small amount of gas). They are NOT gas dwarfs ($H_2$/He-dominated envelope directly on top of a rocky core), because according to condensation/evaporation calculations, there is no way to deplete the less volatile O, N, C more than $H_2$ and He, in fact they should be enriched. For example, in our solar system, the Carbon in Uranus and Neptune [5] is enriched ~50 times solar, and the O, N, C on Jupiter and Saturn are enriched a few up to ~10 times solar [5]. In three aspects: (1) volatility (characterized by equilibrium condensation temperature in the nebula), (2) density, and (3) cosmic abundance, the refractory elements making up silicates and metals (Si, Fe, etc.), the H-compound forming elements (O, N, C), and $H_2$/He always form a ladder or hierarchy, with the elements O, N, C always falling in the middle of this hierarchy:

**Table 1. Hierachy of Planet-Building Elements**

| | Cosmic Abundance (by mass) | Condensation Temperature (K) | Density (solid) (g/cc) |
|---|---|---|---|
| $H_2/He$ | 1000 | 1~10 | 0.2 |
| O, N, C | 6+1+3 | 100~300 | 2 |
| Mg-Silicates | 2 | 1300~1400 | 4 |
| Fe, Ni metal | 1 | 1300~1400 | 8 |

There seems to be a problem with water line, as to whether the water worlds could form so close to the star. However, recent isotope dating of meteorites suggests that planetesimals form quickly after the first condensates from the nebula (<~1 million years), thus allowing wet accretion directly from the nebula before its dissipation [6].

*Alternative View: photo-evaporation*? An alternative view is that the two populations (bi-modal distribution) is a result of photo-evaporation, but there are controversial results in the literature regarding XUV evaporation models, some predict a bimodal distribution [7], while some do not [8].

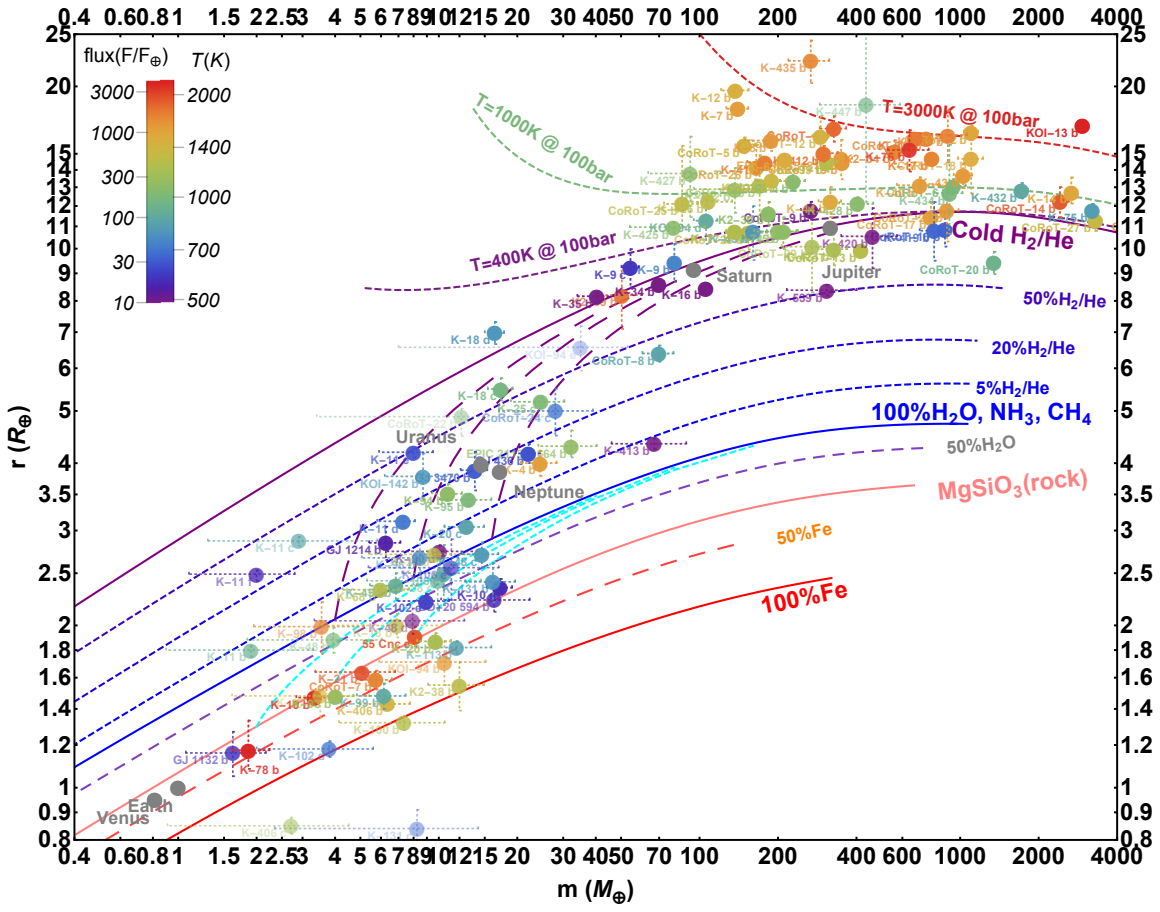

**Figure 3.** Mass-radius diagram of planets with RV-determined masses, color-coded by $T_{equilibrium}$.

**"Main-Sequence" of Planets:** If we restrict our samples to the planets with masses determined by the radial-velocity (RV) method (generally more robust than other methods [9]), then the distribution of planets on the mass-radius diagram form a main-sequence, that is a limited width array, similar to that of stars on Hertzsprung–Russell Diagram (Luminosity-Temperature Diagram). This is consistent with the picture of a material-limited, stepwise growth process, including two kinds of planet building materials – rocky/metallic component and ices.

In **Figure 3**, comparing the positions of planets to the mass-radius curves of various theoretical compositions (color curves as labelled): the first group of planets (<2 $R_\oplus$, red and yellow ones in the lower-left part) complies very well with that of silicates/metal (2:1 weight ratio) composition trend within uncertainty. This trend was first pointed out by [10] and later followed up by [11].

The second group (>2 $R_\oplus$, blue and green ones), on the other hand, complies reasonably well with the following formation scenario: Stage (1) accrete metal/silicate materials up to a few $M_\oplus$. Stage (2) accrete icy materials ($H_2O$, $NH_3$, $CH_4$, i.e., Hydrogen-compounds), if they are available, for another few $M_\oplus$, following the trend of the dashed cyan curves, asymptotically approaching 100% $H_2O$, $NH_3$, $CH_4$ curve. Stage (3) then accrete $H_2/He$ gas, if they are available, following the trend of the dashed purple curves, asymptotically approaching 100% cold $H_2/He$ curve.

*Other evidence.* Contaminated white-dwarf spectra show $H_2O$-rich planet debris. Although, the amount of debris is small compared to 1 Earth mass, it could indicate the existence of water worlds out there, if those debris come from the breakup of those planets [12, 13].

**Conclusion:** The planet formation process leads to three types of planets: rocky, water and gas worlds. They give well-defined fields in the mass-radius diagram as shown in **Figure 4**.

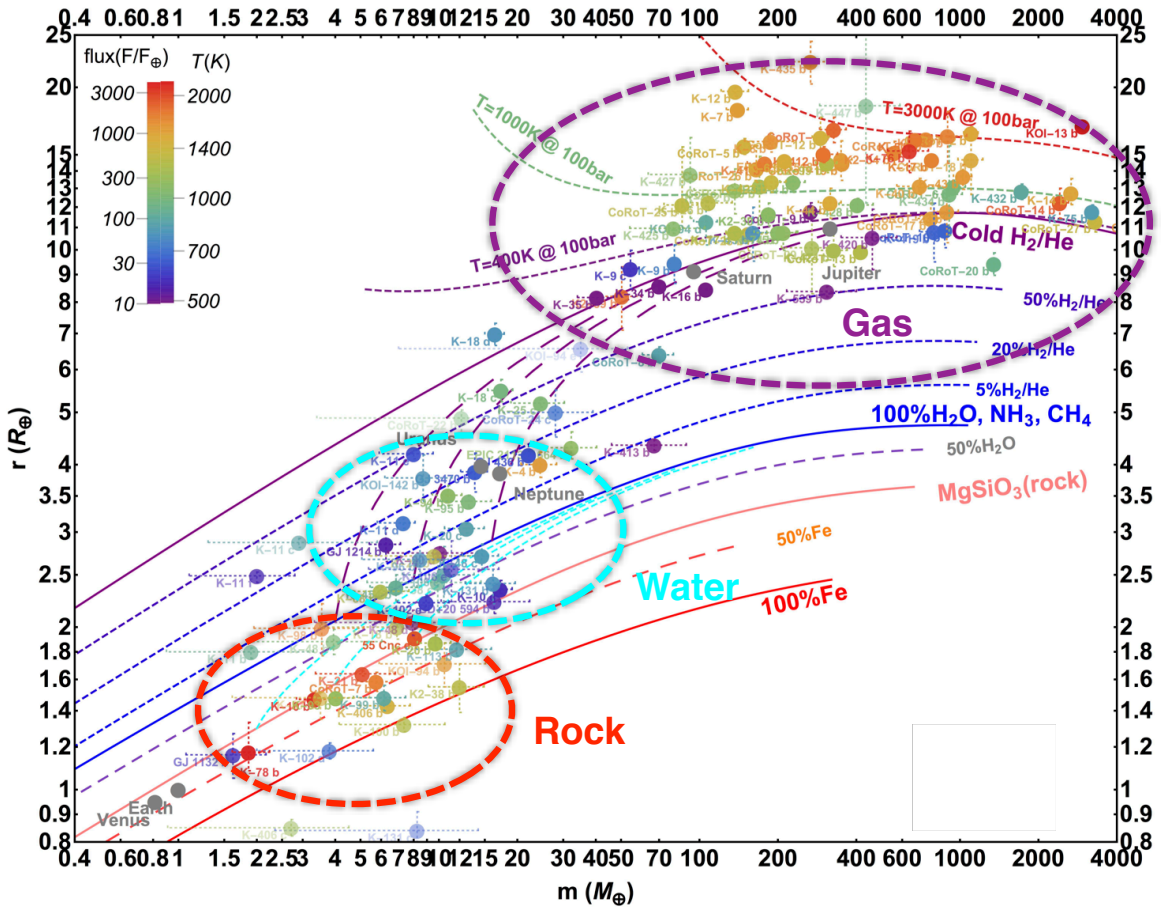

**Figure 4.** Mass-radius diagram of the three types of planets that form in protoplanetary disks.

**References:** [1] NASA Exoplanet Archive. [2] Limpert E. et al. (2001) *BioScience, 51,* 5. [3] Petigura E. A. (2015) *arXiv*:1510.03902 [4] Newton E. R. et al. (2015) *ApJ*, *800*, 85. [5] Atreya S. K. et al. (2016) *arXiv*:1606.04510 [6] Kleine et al. (2009) *GCA*, 73, 5150-5188. [7] Owen J. E. and Wu Y. (2013) *ApJ, 775,* 105. [8] Lopez E. D. and Fortney J. J. (2013) *ApJ, 776,* 2. [9] Hadden S. and Lithwick Y. (2016) *ApJ, 828,* 44. [10] Dressing C. D. et al. (2015) *ApJ, 800,* 135. [11] Zeng L. et al. (2016) *ApJ, 819,* 127 [12] Farihi J. et al. (2013) *Science, 342,* 218. [13] Farihi J. et al. (2016) *MNRAS, 463,* 3186-3192.